\documentclass[journal,10pt,letterpaper]{IEEEtran}

\usepackage{graphicx}
\graphicspath{{Figures/}}
\DeclareGraphicsExtensions{.eps,.ps}
\usepackage{amsmath}
\usepackage{amssymb}

\begin{document}

\title{Annular Spin-Transfer Memory Element}

\author{A.~D.~Kent, {\it Member, IEEE} and D.~L.~Stein%
\thanks{\textit{This work has been submitted to the IEEE for possible
publication. Copyright may be transferred without notice,
after which this version will be superseded.}}
\thanks{Manuscript submitted Januaury 2009}%
\thanks{A.~D.~Kent is with the Department of Physics, New York University, 
  New York, NY 10003 (email: andy.kent@nyu.edu)}
\thanks{D.~L.~Stein is with the Department of Physics and 
  Courant Institute of Mathematical Sciences, New York University, 
  New York, NY 10003 (email: daniel.stein@nyu.edu)}}


\maketitle

\begin{abstract}
An annular magnetic memory that uses a spin-polarized current to switch the magnetization direction or helicity of a magnetic region is proposed. The device has magnetic materials in the shape of a ring (1 to 5 nm in thickness, 20 to 250 nm in mean radius and 8 to 100 nm in width), comprising a reference magnetic layer with a fixed magnetic helicity and a free magnetic layer with a changeable magnetic helicity. These are separated by a thin non-magnetic layer. Information is written using a current flowing perpendicular to the layers, inducing a spin-transfer torque that alters the magnetic state of the free layer. The resistance, which depends on the magnetic state of the device, is used to read out the stored information. This device offers several important advantages compared to conventional spin-transfer magnetic random access memory (MRAM) devices. First, the ring geometry offers stable magnetization states, which are, nonetheless, easily altered with short current pulses. Second, the ring geometry naturally solves a major challenge of spin-transfer devices: writing requires relatively high currents and a low impedance circuit, whereas readout demands a larger impedance and magnetoresistance. The annular device accommodates these conflicting requirements by performing reading and writing operations at separate read and write contacts placed at different locations on the ring. 
\end{abstract}
%

\section{Background}
\label{sec:background}

Nanometer scale magnetic structures enable ultra-fast, light, low power, non-volatile, radiation-hard and inexpensive memory devices that are superior in many ways to conventional random access memories (RAMs), which use electrical charge to store data. In contrast, magnetic RAMs (MRAMs) use the magnetic dipole orientation in small, single domain magnets. Unlike electrical charge, this polarization cannot be ``drained'' away, so the data storage is non-volatile and does not need to be periodically refreshed, leading to superior energy efficiency. Further, in MRAMs data can be accessed rapidly, eliminating the delays inherent to magnetic hard drives and enabling Òinstant startupÓ of computers and other electronic devices.  

MRAM technology has been demonstrated and initial products are on the market \cite{Engel2005}. However, several features of present MRAMs limit their memory density, speed, and reliability. Their geometry decreases stability of the stored information: in particular, the magnetization orientation is susceptible to unpredictable reversal due to random thermal noise. Also, conventional MRAMs rely on the careful application of external magnetic fields to change magnetic polarization (and thereby write information). These applied magnetic fields invariably spread out in space, limiting not only the storage density of conventional MRAM but also their ability to operate efficiently at the nanometer scale.  Some recent devices have addressed this latter problem by using spin transfer rather than applied magnetic fields \cite{Slonczewski1996}, but the problem of stability remains.

The design of the proposed device addresses simultaneously the opposing problems of stable information storage and easy manipulation of information. The former is addressed by utilization of a ring geometry, which calculations have shown results in a far more stable magnetization orientation than other currently used geometries \cite{Martens2006}.  We note that the annular geometry has been proposed in the past for conventional (magnetic field switched) MRAM \cite{Bussmann2001,Moneck2006}.
The latter is addressed through the use of spin current exchange between the two magnetic layers for magnetization switching and excitation. 

The main advantage of using spin transfer to store and extract information is that it allows the magnetic state in the ring to be altered by current pulses applied perpendicular to the layer planes. The pulses transport spin angular momentum between the layers, initiating a change in magnetic state, directly in the region in which the current flows. A measurement of the resistance enables readout of the magnetic state of the ring.  This design avoids the problems of magnetic field spreading, leading to superior speed of writing and readout along with reduction of error due to stray or poorly controlled fields. 

A problem with currently available MRAMs is that writing information typically requires relatively high currents, whereas readout is done with small currents.  The former necessities a low device impedance while the latter requires a high impedance for large readout signal. This leads to a second advantage of the proposed design, which accommodates these conflicting requirements by performing reading and writing operations in different locations, through contacts to different parts of the ring. For example, the readout contact can be through an insulator which forms a magnetic tunnel junction with the ring, while the writing can be through a direct metallic contact to the ring. 

The proposed design carries several other advantages. Present devices often require an optimal size for efficient operation, leaving little tolerance for normal size variation in device fabrication.  However, the magnetization reversal mechanism in the ring geometry is only weakly dependent on ring diameter beyond a small critical size (tens of nanometers)  \cite{Martens2006}. This relative insensitivity to size may lead to greater range of potential use and decreased production costs.

Multielement magnetic devices of the kind proposed here usually have strong magnetostatic interactions between the different elements, which can be hard to quantify or control and therefore lead to problems of density or performance. These interactions are minimized in our design, as magnetic flux is largely confined to the ring.
 
\begin{figure}[b]
  \centering
  \includegraphics[width=0.6\columnwidth]{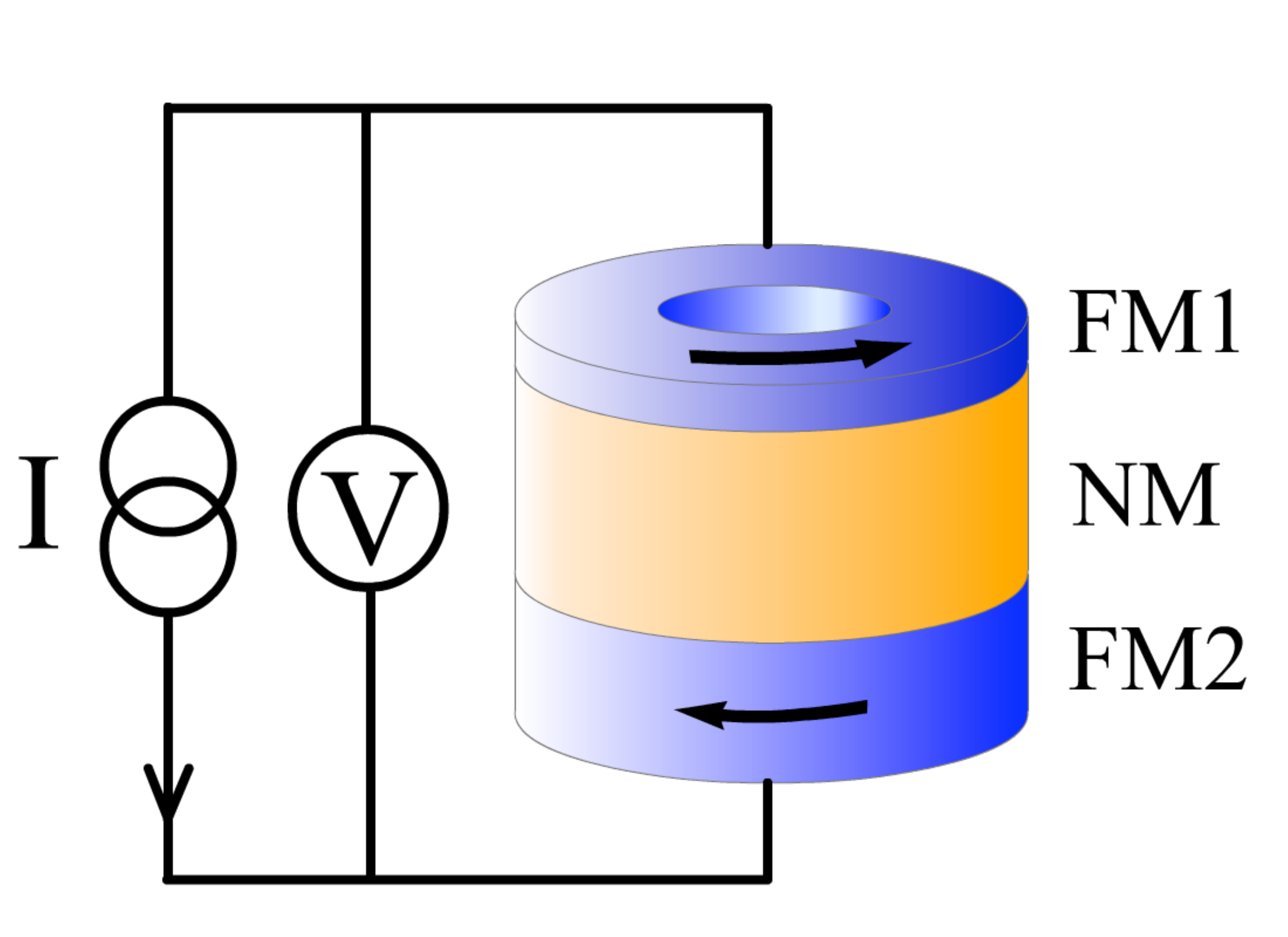}
  \caption{Diagram of 2-terminal annular spin-transfer memory element.}
  \label{fig:diagram1}
\end{figure}

\section{Description of the device}
\label{sec:description}
The proposed device consists of a soft magnetic material (such as a transition metal alloy) in the shape of a thin ring, with thickness of order 1 to 5 nm, mean radius of 20 to 250 nm and ring width between 8 and 100 nm. Its multiple magnetic states (i.e., clockwise versus counterclockwise circulation of the magnetization) enables it to serve as the information storage element. The ring rests on a thin non-magnetic layer, which in turn is supported by a second magnetic layer with a well-defined magnetic state. The non-magnetic layer may be either a non-magnetic metal, such as Cu, or a thin insulating layer. In the former case, the device is a giant magnetoresistance multilayer stack. In the latter case, the device contains a magnetic tunnel junction in which electrons traverse the insulating layer by tunneling. This bottom layer should be magnetically harder than the ring, either through greater thickness or through different material composition. Possibilities for the latter include use of a material with  larger magnetic anisotropy (e.g., cobalt), or through exchange coupling to an antiferromagnetic layer (e.g., IrMn or FeMn) \cite{Ross2006}. Possible materials for the free magnetic layer include  NiFe (permalloy) and CoFeB. Electrical contacts are placed at the top and bottom of the structure, so that current can be injected and flow perpendicular to the plane of the ring. 

\begin{figure}[t]
  \centering
  \includegraphics[width=0.95\columnwidth]{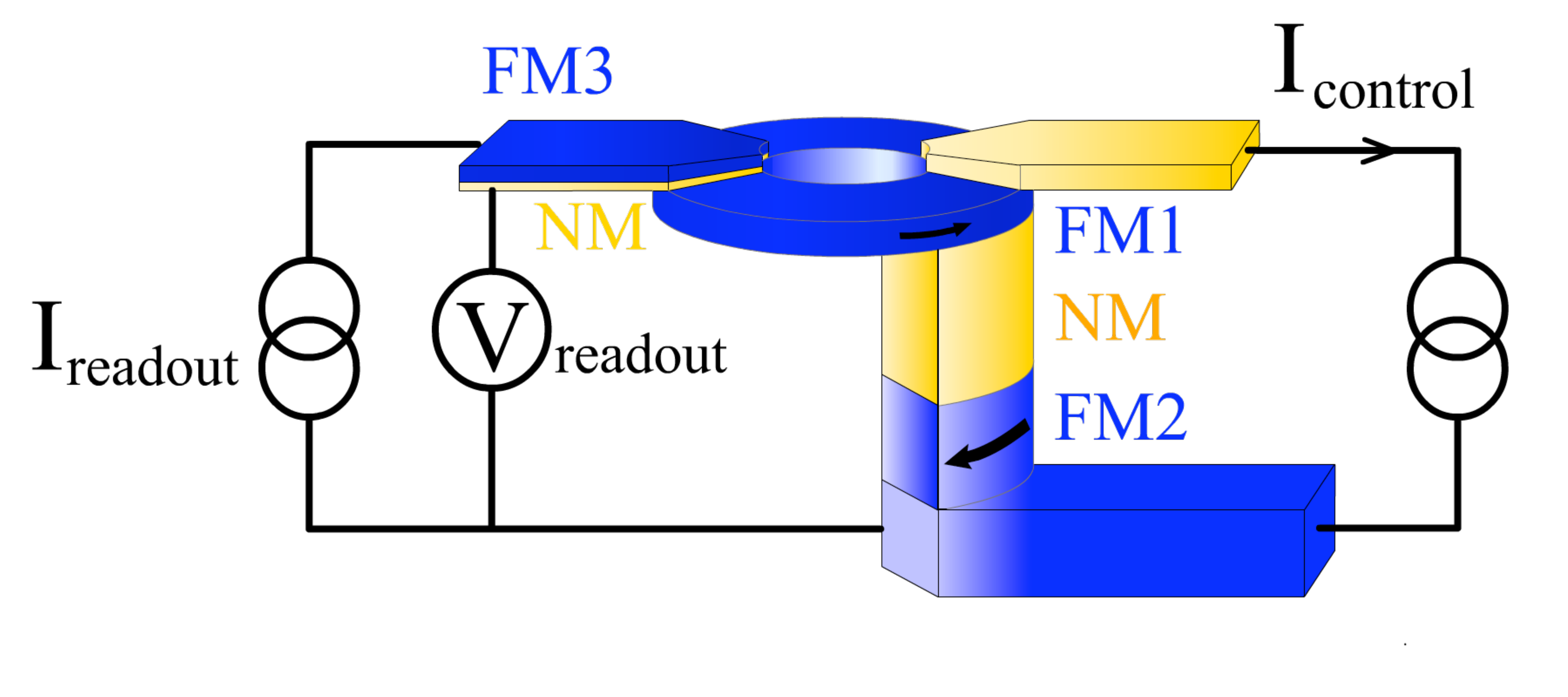}
  \caption{Diagram of 3-terminal device with separate read and write contacts, permitting independent optimization of the read and write circuitry.}
  \label{fig:diagram2}
\end{figure}

The basic device geometry is shown in Fig.~\ref{fig:diagram1}. The resistance of the device depends on the relative magnetization directions of the free and fixed magnetic annuli due to the giant magnetoresistance (NM=nonmagnetic metal) or tunnel magnetoresistance effects (NM=thin insulator). Typically, the state with opposite magnetization helicity of the free and fixed magnetic layers will have a larger resistance. A small read current is employed to determine the element resistance, and thereby, the information stored. Larger current pulses are used to switch the helicity direction of the free magnetic layer, change the resistance state of the device and write information. In this geometry the circuit for reading and writing information is the same, that is, this is a two terminal device. However, as discussed above, it may be advantageous to have separate read and write circuits that can be optimized independently. Fig.~\ref{fig:diagram2} is an example of such a device. 

Fig.~\ref{fig:diagram2} illustrates a 3-terminal annular spin transfer memory device. The memory state is written by sending a control current ($I_{\sf control})$ through a low impedance contact to a portion of the ring. The readout contact forms a magnetic tunnel junction with the free layer. It consists of a thin insulating layer between the free layer and the FM contact (FM3) with a fixed magnetization direction. Electrons traverse this layer by tunneling, which produces a larger impedance, readout signal and magnetoresistance \cite{Moodera1995,Parkin2004,Yuasa2004,Dave2006}. Therefore a small readout current may be employed, well below that needed to initiate magnetization reversal. FM3's magnetization direction is fixed to maximize the resistance change associated with changes in the magnetization helicity of the ring, i.e., its magnetization direction is tangential to the ring circumference.

\subsection{Stability of helical magnetization states}

The magnetization of the device must be stable over long time periods under static operating conditions and yet easily changed or reversed under the action of a current pulse. Here we discuss the stability of helical magnetization states. Generally, the advantages of this device in terms of long term storage is that an annular shape minimizes the number of edges or corners that may act as nucleation sites for undesired magnetization reversal--which reduces the rate of such events. Helical magnetization states are also extremely stable, because the magnetostatic energy is minimized. Additionally, the magnetization reversal mechanism of ring geometries is only weakly dependent on ring diameter beyond a critical size \cite{Martens2006}, which may enable greater fabrication tolerances.

The energy barrier, transition states and rates of thermally induced reversal have recently been found for ferromagnetic rings, both analytically in a one-dimensional approximation \cite{Martens2006} and through numerical micromagnetic studies \cite{Chaves2008,Chaves2008a}. The 1-D analytic model has been shown to apply over a wide range of ring dimensions, encompassing those of interest for an annular spin-transfer device composed of transition metal ferromagnets. Importantly, it accurately estimates the energy barrier and transition states, as verified by numerical simulations \cite{Chaves2008,Chaves2008a}.

The rate $\Gamma$ of thermally induced transitions between two minima in the limit of low noise is given by the Arrhenius formula
\begin{equation}
\label{eq:Kramers}
\Gamma \sim \Gamma_0 \exp(-U/k_B T),
\end{equation}
where $U$ is the energy barrier, $k_B$ is Boltzmann's constant, and
$T$ is the temperature.  The rate prefactor $\Gamma_0$ was calculated explicitly in Ref.~\cite{Martens2006} and is of order inverse ferromagnetic resonance frequency ($\sim 10^{-9}$ s). In order to minimize undesired thermally induced reversal ($1/\Gamma \gg 10$ years), $U \ge 60 k_B T$ is required.

\begin{figure}[t]
  \centering
  \includegraphics[width=0.95\columnwidth]{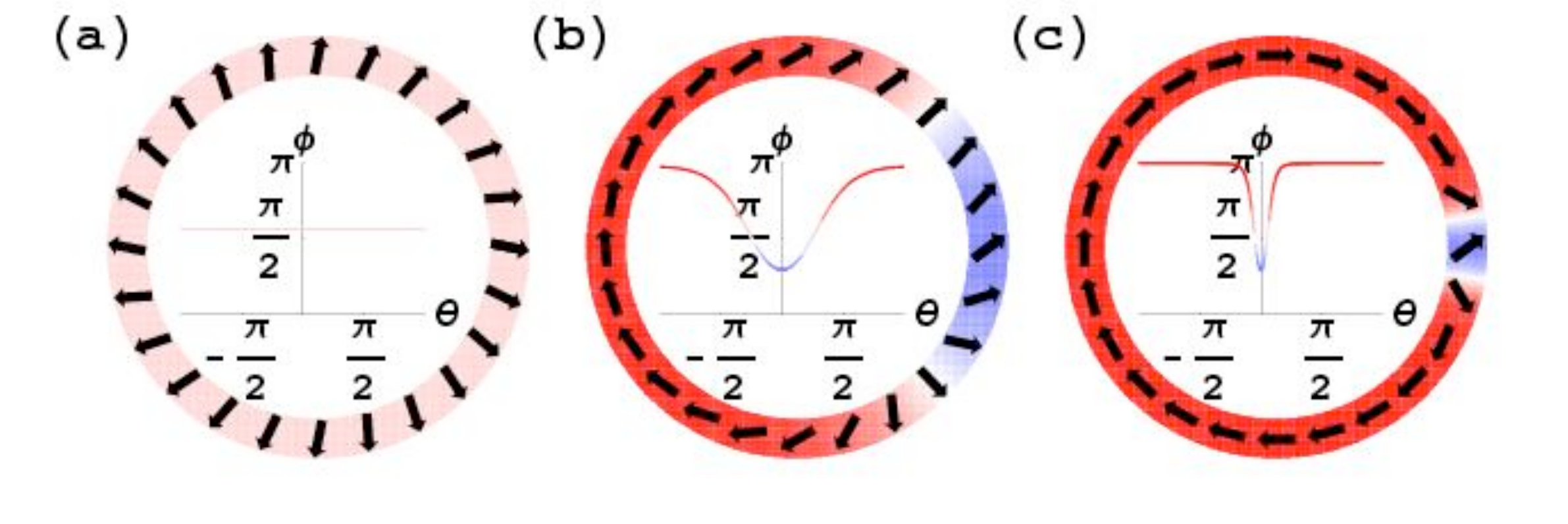}
  \caption{Saddle configurations for $h=0.2$. (a) Constant saddle, (b) instanton saddle for $\ell=12$ and (c)
  instanton saddle for $\ell=60$. From ref. \cite{Chaves2008}}
  \label{fig:saddles}
\end{figure}

The model of Ref.~\cite{Martens2006} determined the energy barrier to reversal as a function of material parameters, ring dimensions and applied (circumferential) magnetic field. Key parameters are the normalized field and ring size:

\begin{eqnarray}
\label{eq:scaled1}
h=\frac{H_e}{H_c}=\frac{H_e}{\frac{M_o}{\pi}\left(\frac{t}{
\Delta R} \right)\left | \hbox{ln}\left(\frac{t}{R}\right)\right |}\\
\ell=\frac{R}{\lambda}\sqrt{2\pi \left(\frac{t}{ \Delta R}\right)
\left|\hbox{ln}\left(\frac{t}{R}\right)\right|}\, .
\end{eqnarray} 
Here $M_o$ is the saturation magnetization, $t$ is the ring thickness,
$\Delta R$ is the ring width, $R$ is the average radius, $H_e$ is the
external magnetic field, and $H_c$ is the field at which the metastable
configuration becomes unstable. Note that $h$
also represents the ratio of the Zeeman energy to the (easy plane) shape
anisotropy energy~\cite{Martens2006}.  The exchange length is given by
$\lambda=\sqrt{2A/(\mu_{0}M_{o}^{2})}$, where $A$ is the exchange constant. 
$\ell$ represents the ratio ring size to the width of a Bloch wall. 
For $\ell \le 2\pi\sqrt{1-h^2}$ the theory predicts a constant saddle, as
shown in Fig.~\ref{fig:saddles}a, whereas for $\ell>2\pi\sqrt{1-h^2}$, it
predicts an instanton saddle (Figs.~\ref{fig:saddles}b, c). Both of these
saddle configurations are described by a function $\phi_{h,l}(\theta)$ 
~\cite{Martens2006}.

\begin{figure}[b]
  \centering
  \includegraphics[width=0.95\columnwidth]{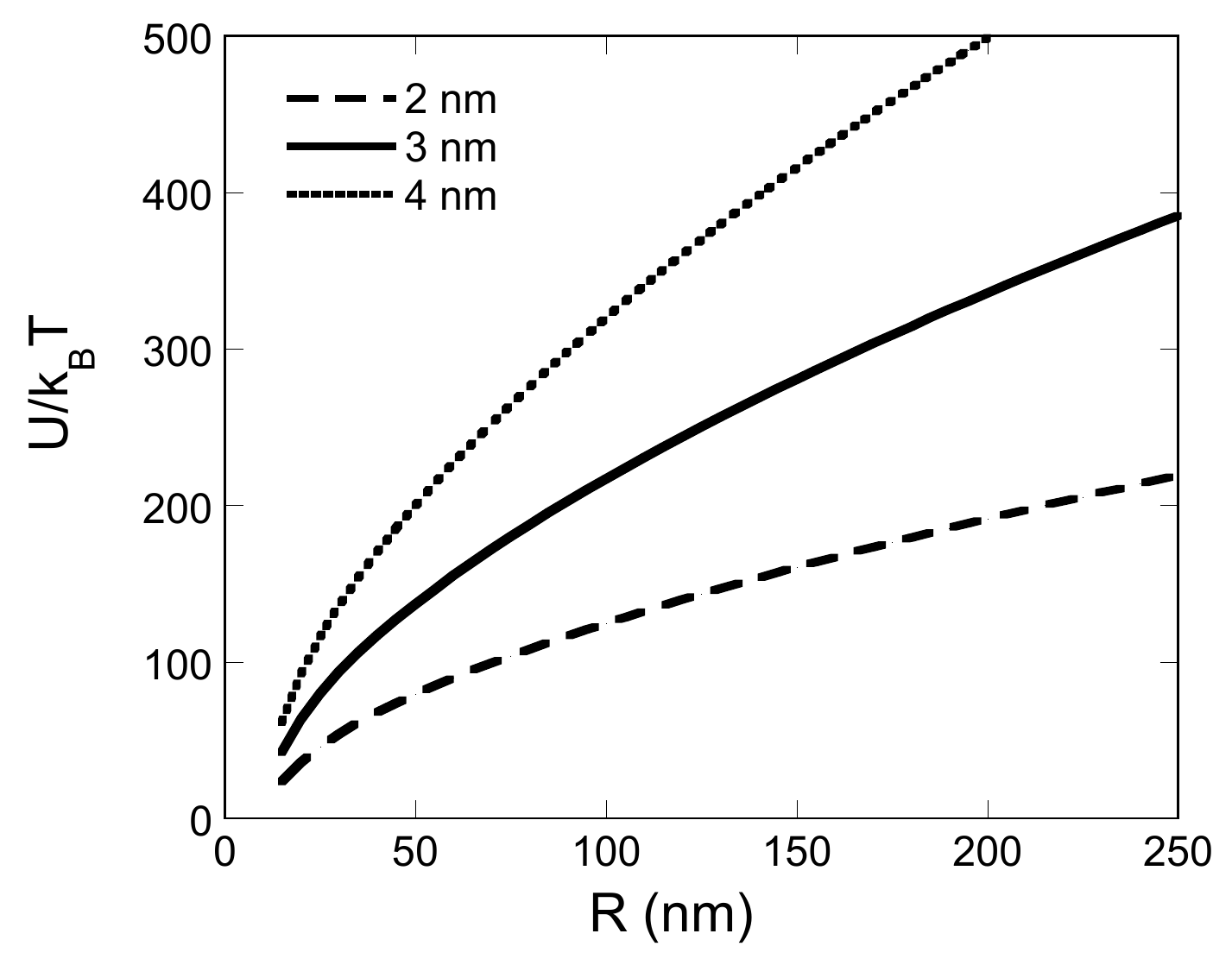}
  \caption{Energy barrier to magnetization reversal for permalloy rings as a function of mean radius in zero applied field. The ring width is $\Delta R=0.4R$ and ring thickness is $2$, $3$ and $4$ nm. The energy barrier is plotted in units of $k_BT$ for $T=300$ K.}
  \label{fig:EnergyBarriers}
\end{figure}

The scale of the energy barrier is given by:
\begin{equation}
\label{eq:Energyscale}
E_0= {\mu_o M_o^2 \over \pi} {\Delta R \over R}  \ell t \lambda^2
\end{equation}
For the constant saddle $U=E_0 (1-h)^2 \ell /2={\mu_o M_o^2} t^2 R |\ln
t/R| (1-h)^2$. This is independent of the exchange length, since the
transition state has magnetization at a constant angle to the ring
circumference (Fig.~\ref{fig:saddles}a). For the instanton saddle the
result is in general more complicated (cf. Eq. 13 of
~\cite{Martens2006}). However, in the limit $\ell \gg 2 \pi$ the energy
barrier is  $U = 4E_0(\sqrt{1-h}-h\sec^{-1}{\sqrt{h}})$. This can easily be
greater than $60 k_B T$ at room temperature for rings fashioned from
permalloy or CoFeB. For example, a permalloy ring ($A=1.3 \times 10^{-11}$
J/m and $M_o=8 \times 10^5$ A/m) with $R=50$ nm, $\Delta R=20$ nm and $t=2$
nm is in the large $\ell$ limit ($\ell=12.6$) and the energy barrier
associated with the instanton saddle is $U/k_B(300$ K$)=80$ at $h=0$. Larger
rings sizes with $\Delta R \lesssim R$ have greater energy barriers to
reversal and therefore easily satisfy the requirement of long term
stability for data retention. Figure~\ref{fig:EnergyBarriers} shows the
dependence of the energy barrier to magnetization reversal on ring radius
for difference ring thicknesses ($t=2$, $3$ and $4$ nm).

\subsection{Current induced switching}

A current pulse transports spin-angular momentum that can change the helicity of the thin free magnetic layer. The Landau-Lifshitz-Gilbert equations of motion for the magnetization of the free layer, $\bf{M}$, including a spin-current induced torque is: 
\begin{eqnarray}
{{\partial {\bf M}} \over{\partial t} }=-\gamma \mu_o{\bf M} \times {\bf H_\text{eff}} +{\alpha \over M_o} {\bf M}
\times {{\partial {\bf M}} \over{\partial t}}\nonumber\\ + {\gamma a_{J} \over M_o}
{\bf M} \times ({\bf M} \times \hat{m}_p).
\label{eq1}
\end{eqnarray}
$\hat{m}_p$ is a unit vector in the direction of
magnetization of the fixed magnetic layer and
$\gamma$ is the gyromagnetic ratio, $\gamma=|g\mu_B/\hbar|$. The second term on the right is
the damping term, where $\alpha$ is the Gilbert damping constant.
The last term is due to spin-transfer. 
The vector cross product, ${\bf M} \times ({\bf M} \times \hat{m}_p)$, is in the direction of spin
angular momentum transverse to the free layer magnetization and in the plane containing 
${\bf M}$ and $\hat{m}_p$. 

The prefactor of the spin-transfer term, $a_{J}$, depends on the
current, $I$ and the spin-polarization of the current, $P$, $a_{J} = \hbar P I/(2e\mu_oM_oV)$
 \cite{Slonczewski1996}, where $V$ is the volume of the magnetic element. 
We define positive current to correspond to the flow of electrons from the thin free ferromagnetic layer (FM1)
to the pinned ferromagnetic layer (FM2).

Generally positive current favors anti-alignment of the layer magnetizations
and a high resistance state, with the reference and free layers having
opposite helicity (an AP state). Negative current favors a state with the
same helicity (a P state). The fixed layer magnetization can be set such
that the spin-transfer interaction and the current induced Oersted fields,
(the circumferential) fields generated by the current, favor the same
configurations. For instance, this requires that the fixed layer
have a clockwise magnetization configuration, as shown in Figs.~\ref{fig:diagram1}~and~\ref{fig:diagram2}.

We now calculate the spin-transfer current density needed to destabilize
the P and AP states.  The P state is
given by $\hat{m}=-\hat\theta$ and the AP state by $\hat{m}=+\hat\theta$ (Fig.~\ref{fig:diagram1}) where
$\hat{m}_P=-\hat\theta$. Also, the applied field associated with the current flow through the ring (along $z$) is ${\bf H_e}=H_e\hat\theta=I\hat\theta/(2\pi R)$. The effective field $\mu_o{\bf H_{\rm eff}}=-\delta
E/\delta{\bf{M}}$ is the variational derivative of the total energy density $E$,
given by~\cite{Martens2006} 
\begin{eqnarray}
\label{eq:energy}
E={E_0 \ell \over 4 \pi} \int_0^{2\pi}d\theta \Big[ \Big({\partial
m_r\over\partial\theta}\Big)^2+\Big({\partial
m_\theta\over\partial\theta}\Big)^2+\Big({\partial m_z\over\partial\theta}\Big)^2\nonumber\\
+m_r^2-2hm_\theta+dm_z^2\Big]\,
\end{eqnarray}
written in terms of the normalized magnetization vector in cylindrical coordinates, 
${\bf M}/M_o=(m_r,m_\theta,m_z)$. $d$ is the ratio of the
out-of-plane anisotropy to the in-plane anisotropy, $d=2\pi^2R^2/(\ell^2\lambda^2)$ and is typically much larger
than $1$. For example, a permalloy ring with $R=50$ nm, $\Delta R=20$ nm and $t=2$ nm has $d=10$.

A straightforward linear stability analysis using the LLG equations then
give for the P state with both the reference and free layers with clockwise magnetization configurations
($m_\theta=-\hat\theta$):
\begin{equation}
\label{Pinstability}
a_J < {E_0 \ell \alpha \over 2 \mu_o M_oV}(d+1-2h).
\end{equation}
The Oersted field generated by the current ($h$) renders the clockwise configuration of the free
layer metastable and lowers the threshold current for spin-torque induced switching. In the AP state the free layer is magnetized counterclockwise and the stability condition is:
\begin{equation}
\label{APinstability}
a_J > -{E_0 \ell \alpha \over 2 \mu_o M_oV}(d+1+2h).
\end{equation}
Negative currents (less than $a_J$) lead to switching from the AP to P states. In this case, the Oersted fields are reversed ($h<0$) and again this field lowers the magnitude of the current needed for spin-torque induced switching.

We note that new physics enters when the Oersted field approaches the value needed to render the metastable state unstable, i.e., when $h\rightarrow1$.
There is a critical field, $h_c$, less than 1, above which the stability conditions above change: 
\begin{equation}
h_c = 1/2(\sqrt{1+\alpha^2}+1)-d/2(\sqrt{1+\alpha^2}-1),
\end{equation}
or for $\alpha^2 \ll d\alpha^2\ll 1$
\begin{equation}
h_c \simeq 1-d\alpha^2/4.
\end{equation}
For $h>h_c$ the stability condition eqn. \ref{Pinstability} is:
\begin{eqnarray}
a_J  < {E_0 \ell \alpha \over 2\mu_o M_oV} \Big( d + 1 - 2h \nonumber\\ - 1/\alpha\sqrt{\alpha^2(d-1)^2 - 4(d-h)(1-h)}\Big).
\end{eqnarray}
Note that in the limit in which the metastable state becomes unstable ($h\rightarrow1$) the threshold current goes to zero, $a_J\rightarrow0$.

It is interesting to estimate the current threshold for spin-transfer switching. We neglect the Oersted field, i.e., we take $h=0$. The threshold current is given by
\begin{equation}
\label{eq:threshold}
I_T= {e \alpha E_0 \ell \over  \hbar P}(d+1).
\end{equation}
We consider a permalloy ring with $R=50$ nm, $\Delta R=20$ nm and $t=2$ nm. With $\alpha=0.01$ and a spin-polarization, $P=0.4$ we find that $I_T=440 \;\mu$A and a current density $J_T=6 \times 10^6$ A/cm$^2$. Note that this current produces a circular Oersted field of $H_e=1350$ A/m, corresponding to $h=0.02$. So the Oersted field is negligible and the spin-torque interaction is far more effective at switching the magnetization direction than magnetic fields associated with the current. 

\subsection{Spin-Transfer Figure of Merit}
A figure of merit for spin-transfer devices is the ratio of the threshold current to the energy barrier for reversal:
\begin{equation}
\epsilon=I_T/U.
\end{equation}
The smaller $\epsilon$ the better the device performance; the smaller the current required to switch the device and the lower the energy dissipation in writing information. The energy barrier for the constant saddle configuration with $h=0$ is $U=E_0 \ell/2$, which gives:
\begin{equation}
\epsilon={2e \alpha \over  \hbar P} (d+1).
\label{eq:epsilon}
\end{equation}
 \begin{figure}[b]
  \centering
  \includegraphics[width=0.95\columnwidth]{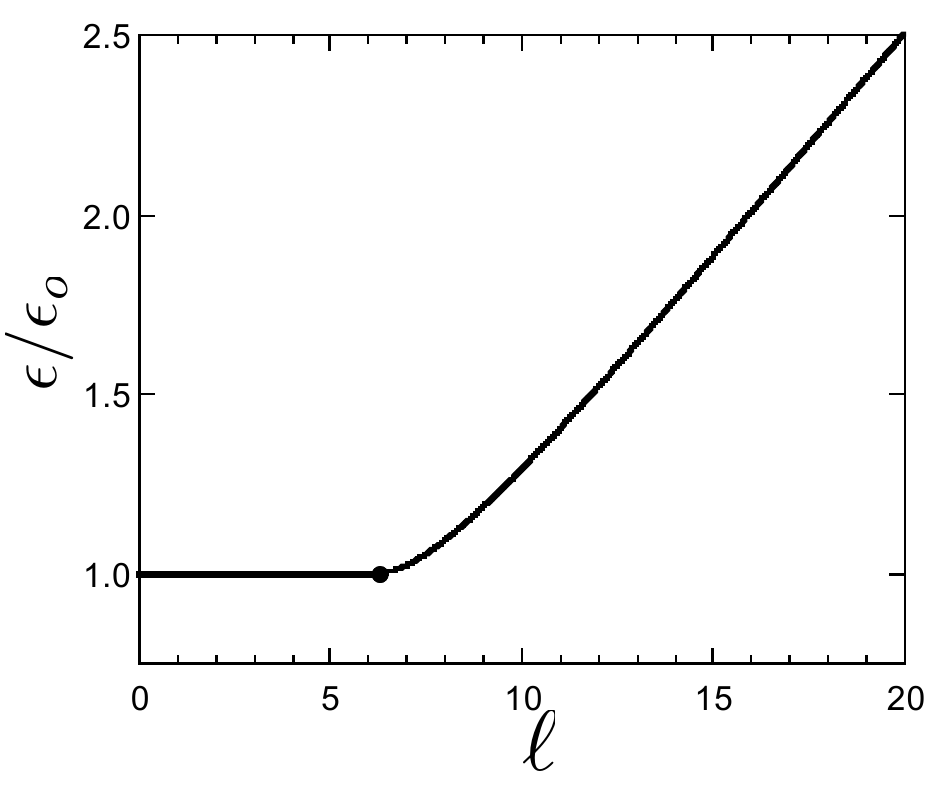}
  \caption{Normalized spin-transfer figure of merit, $I_T/U$, versus ring size, $\ell$, where $\epsilon_o=2e \alpha(d+1)/ (\hbar P)$ is figure of merit in the constant saddle regime, $\ell < 2\pi$.  The dot indicates the transition from the constant to instanton transition state.}
  \label{fig:eff}
\end{figure}
Reducing the damping and/or increasing the spin-polarization of the current leads to lower current thresholds and a more energy efficient device. For comparison, the smallest figure of merit has been found in perpendicularly magnetized thin film nanoelements \cite{Sun2000,Mangin2006}. These are nanopillars in which both the free and fixed layer are magnetized perpendicular to the plane of the element, along $z$. In the macrospin limit the energy of the free layer is given by $E=-Um_z^2$ and $\epsilon=4e \alpha /(\hbar P)$.

As the ring size increases the instanton becomes the preferred saddle configuration and the figure of merit increases. This is shown in Fig.~\ref{fig:eff}. The dot in Fig.~\ref{fig:eff} marks the transition from the constant to the instanton saddle. In the limit $\ell \gg 2\pi$, $\epsilon/\epsilon_o \rightarrow \ell/8$, where $\epsilon_o=2e \alpha(d+1)/ (\hbar P)$. 

Most spin-transfer MRAM devices consist of planar thin film elements composed of soft magnetic materials (i.e., permalloy) with an asymmetric shape that leads to an easy axis in the plane of the element. The fixed and free layers are magnetized in the plane (along $x$) and the energy of the free layer is $E=U(m_y^2+dm_z^2)$ in a macrospin model. Their spin-transfer figure of merit in this macrospin limit is given by Eq. \ref{eq:epsilon}, i.e. it is the same as that of an annular spin-transfer device in the constant saddle regime ($\ell < 2\pi$). As the element size increases we expect the same qualitative behavior for the figure of merit as that for the ring geometry; the figure of merit is expected to increase with the element size when the macrospin model breaks down and thermally induced magnetization reversal occurs via nonuniform modes.

We note that since $d\gg1$ annular spin-transfer and planar thin film elements are less efficient than perpendicularly magnetized thin film nanoelements. However, an advantage of soft magnetic materials is there relative low Gilbert damping parameters ($\alpha < 0.01$). Recent research has shown that materials with perpendicular magnetic anisotropy have larger Gilbert damping, $\alpha \simeq 0.04$ \cite{Beaujour2007,Chen2008}. However, fully spin-polarized materials have recently been shown to have record low Gilbert damping parameters ($\alpha=0.003$) \cite{Heinrich2004} and are thus of great interest for annular spin-transfer devices.
\subsection{CMOS Integration}
\begin{figure}[t]
  \centering
  \includegraphics[width=0.95\columnwidth]{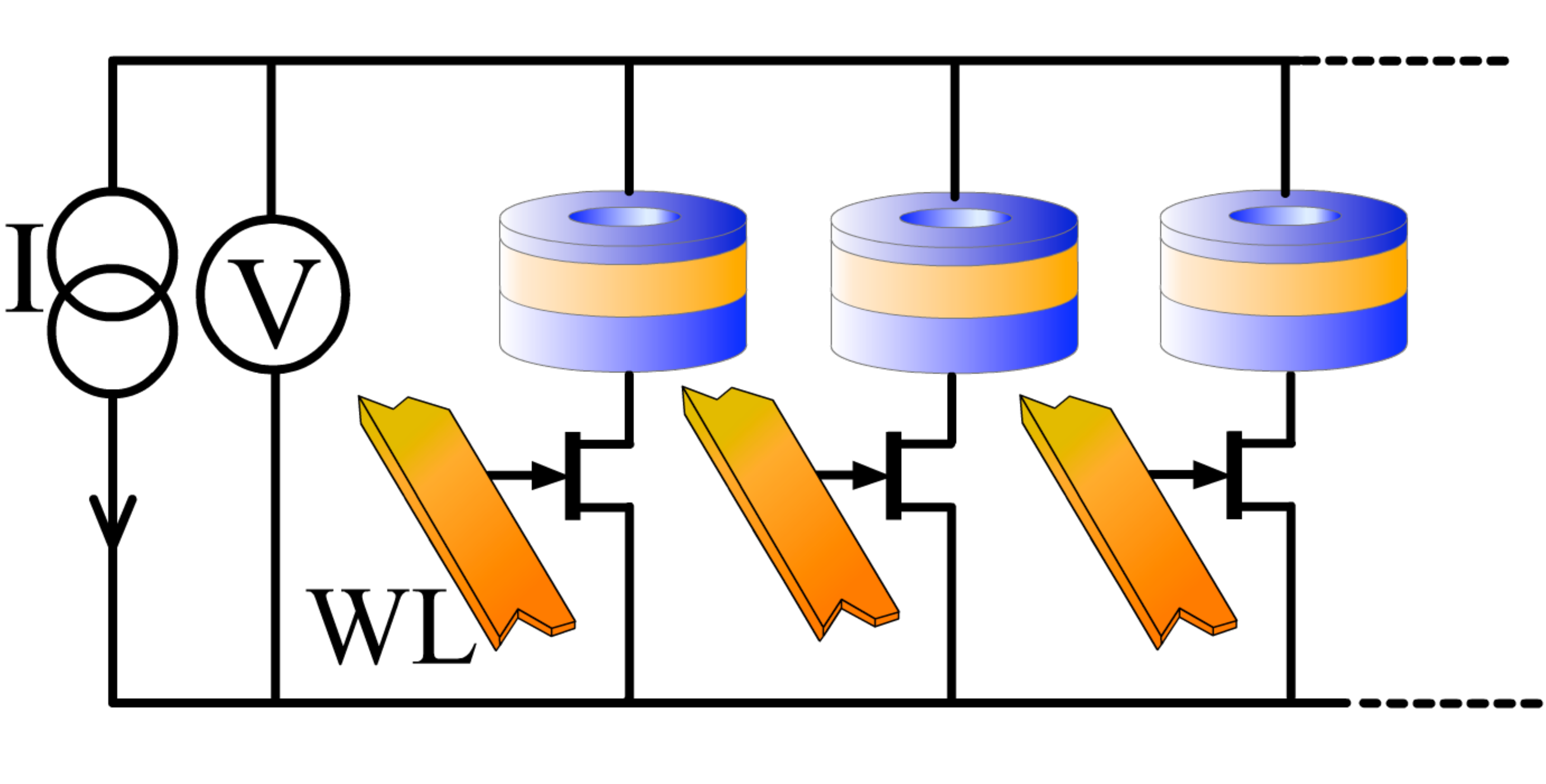}
  \caption{A schematic of the annular spin transfer device memory cell architecture. Each cell has one transistor. A voltage is applied on the word line (WL) to address and activate a particular element in memory array.}
  \label{fig:Circuit}
\end{figure}

A bit cell would consist of magnetic ring and one transistor for current control and readout (Fig.~\ref{fig:Circuit}). The current density per unit gate width is typically $1$ mA/$\mu$m for a CMOS transistors. Therefore smaller switching currents permit smaller minimum feature size, $f$, transistors and larger device integration density. The memory density of an annular device will thus be determined by the ring size and the switching current. Our device composed of permalloy with $R=50$ nm ($\Delta R=20$ nm, $t=2$ nm, $\alpha=0.01$ and $P=0.4$) requires transistors with 0.5 $\mu$m gate length. Assuming a bit lateral size of four times the minimum feature size, $4f$, gives a bit areal density greater than $10^7$ device/cm$^2$. This relatively low integration density could be increased by reducing the device operating currents, for example, by employing materials with smaller damping.
\section{Summary}
Annual spin-transfer devices offer helical magnetization states that are stable at room temperature at device radii less than $50$ nm. They are composed of soft magnetic materials that are widely used in conventional (field-switched) MRAM. The model presented here shows that a spin-polarized current can readily and efficiently switch the magnetization helicity of device to write information. The switching currents are predicted to be low enough to enable the realization of a high density energy efficient MRAM.  Further the ring configuration leads to minimal dipolar interactions between memory elements. We finally note that recent experimental studies of rings have focused on the so-called onion states, where there is a net magnetization along the along a direction in the ring plane that can be altered with uniform magnetic fields or a spin-polarized current \cite{Wen2007}. The helical states of rings are far more stable and thus likely to be of greater utility in miniaturized and high density magnetic memories.

\section*{Acknowledgments}
We acknowledge discussions with Gabriel Chaves and Jean-Marc Beaujour on this project. This work was supported by NSF Grant Nos. DMR-0706322 (ADK) and PHY-0651077
(DLS).
\bibliographystyle{IEEEtran}

\begin{IEEEbiographynophoto}{Andrew D. Kent}
received the Ph.D. degree in applied physics from Stanford University, 
Stanford, CA in 1988. Since 1994, he has been a Professor of Physics at New York
University. His research interests center on the physics of magnetic nanostructures, 
nanomagnetic devices and magnetic information storage. He has conducted experimental studies of quantum tunneling of magnetization and coherence in arrays of nanometer scale magnets known as single molecule magnets. He has also studied spin-dependent transport and spin momentum transfer in thin film magnetic nanopillars.
\end{IEEEbiographynophoto}

\begin{IEEEbiographynophoto}{Daniel L. Stein}
received the Ph.D.\ degree in physics from Princeton University, 
Princeton, New Jersey, in 1979.  Since 2005, he has been a Professor 
of Physics and Mathematics at New York University.  
His research in recent years has focused primarily on quenched disorder 
in statistical mechanical systems, particularly spin glasses, and 
on stochastic processes in both macroscopic and nanoscale systems.
\end{IEEEbiographynophoto}
\vfill

\end{document}